\documentclass[english,aps]{revtex4}
\usepackage[T1]{fontenc}
\usepackage[latin1]{inputenc}
\usepackage{graphicx}
\usepackage{amssymb}

\makeatletter

\providecommand{\tabularnewline}{\\}

\usepackage{graphicx}

\usepackage{graphicx}
\usepackage{calc}
\newlength{\nrWidth}
\usepackage{babel}
\makeatother
\begin{document}

\title{Rashba spin-orbit coupling and spin relaxation in silicon quantum
wells}

\author{Charles Tahan and Robert Joynt}

\affiliation{Physics Department, University of Wisconsin-Madison}

\date{January 2004}

\begin{abstract}
Silicon is a leading candidate material for spin-based devices, and
two-dimensional electron gases (2DEGs) formed in silicon heterostructures
have been proposed for both spin transport and quantum dot quantum
computing applications. The key parameter for these applications is
the spin relaxation time. Here we apply the theory of D'yakonov and
Perel' (DP) to calculate the electron spin resonance linewidth of
a silicon 2DEG due to structural inversion asymmetry for arbitrary
static magnetic field direction at low temperatures. We estimate the
Rashba spin-orbit coupling coefficient in silicon quantum wells and
find the $T_{1}$ and $T_{2}$ times of the spins from this mechanism
as a function of momentum scattering time, magnetic field, and device-specific
parameters. We obtain agreement with existing data for the angular
dependence of the relaxation times and show that the magnitudes are
consistent with the DP mechanism. We suggest how to increase the relaxation
times by appropriate device design. 
\end{abstract}
\maketitle

\section{\textbf{Introduction} \emph{}}

Electron spins in silicon have been proposed as an attractive architecture
for spintronics and quantum information devices. The inherently low
and tunable spin-orbit coupling (SOC) in silicon heterostructures
and the possibility of eliminating hyperfine couplings by isotopic
purification bodes well for quantum coherent spin-based qubits and
spin transport. Early experiments together with theory have shown
that coherence times can be upwards of three orders of magnitude longer
than in GaAs.\cite{ART-Jantsch/Schaffler-2002,ART-Tryshkin/Lyon-ESRof2DEGs-2003,ART-Tryshkin/Lyon-ESRofPSi-2003,ART-deSousa-NuclearDiffusion-2003}

Energy relaxation of localized spin states has attracted theoretical
attention \cite{ART-Roth-gFactorandSpinLatticeRelaxationinGeandSi-1960,ART-Hasegawa-SpinLatticeRelaxationinSiGe-1960,ART-Khaetskii-ZeemanFlipInDots-2001,ART-Pines-SpinRelaxationInSilicon-1957}
and experimental effort \cite{ART-Wilson/Feher-ESR3-1961,ART-Tryshkin/Lyon-ESRofPSi-2003}
for decades, and this activity has recently revived in the context
of quantum computation. The idea is to store quantum information in
the spin of a single electron confined in a semiconductor structure,
either attached to a donor atom or confined electrostatically in a
quantum dot. Spin transport, also of great interest, encodes information
in the spin states of an ensemble of electrons. In both cases, electron
spin resonance (ESR) measurements of spin relaxation provide a key
and available measure of spin coherence properties of electrons in
silicon quantum wells, though not a one-to-one correspondence. Our
aim in this paper is to explain some existing ESR results for silicon
2DEGs at low temperatures and to make predictions for future experiments.

\begin{figure}
\begin{center}\includegraphics[%
  scale=0.4]{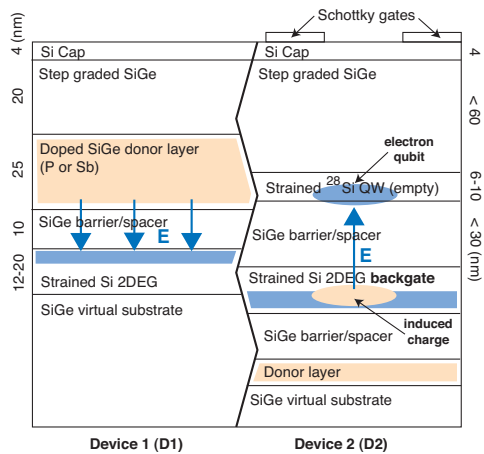}\end{center}

\caption{Strong, internal electric fields are common in silicon quantum well
devices. \emph{D1:} A typical, high-mobiliy SiGe heterostructure uses
a donor layer to populate a high-density 2DEG. The charge separation
results in an $E_{z}\sim10^{6}$ V/m. \emph{D2:} A proposed quantum
dot quantum computer \cite{ART-Friesen-PracticalQDQCinSi-2002} which
utilizes a tunnel-coupled backgate to populate the quantum well without
the need for a nearby donor layer. Here, $E_{z}>10^{5}$ V/m due to
the image potential formed on the backgate.\label{cap:devices}}
\end{figure}

The structures that concern us here are layered semiconductor devices
of Si and SiGe. The active layer is the quantum well (QW) that confines
the electrons in the growth direction. This layer will be assumed
to be composed of pure, {[}001{]} strained silicon. We shall also
neglect any roughness or miscut at the the Si/SiGe interfaces. Devices
made in this way are commonly referred to in the semiconductor industry
as MODFETs and are designed to maximize mobility. Figure \ref{cap:devices}
introduces two example structures.

Extensive theoretical work has been done on spin relaxation in GaAs
and other III-V materials. The developments that began with the theory
of D'yakonov and Perel' \cite{ART-Dyakonov/Perel-3DEG-1972,ART-Dyakonov/Kachorovskii-1986,BOOK-Optical-Orientation}
are most relevant for our purposes. These authors found that fluctuating
effective magnetic fields due to momentum scattering in the presence
of SOC is the dominant spin relaxation mechanism in semiconductor
2DEGs for low temperatures. Here, we start with this assumption and
use a general spin-density matrix approach to calculate the relaxation
times of a 2DEG in the presence of a static magnetic field, including
explicitly the angular dependence. Understanding the angular dependence
of the linewidth is important for comparison with ESR experiments
and the extraction of physically relevant parameters such as the momentum
relaxation time. We also calculate separately in a $k\cdot p$ formalism
the Rashba spin orbit coupling parameter in silicon quantum wells.
This is a key parameter in our calculation as well as for other spin
control considerations, both negative and positive.

In the next section we discuss the origin and magnitude of the SOC
in realistic heterostructures. The section following that presents
our calculation. Lastly, we compare with experiment and discuss the
implications for device design.

\section{\textbf{Spin-orbit coupling}}

The strong macroscopic electric fields inside heterostructure QWs
are important for understanding SOC, especially in silicon. These
fields are also device-specific, so we carry out our calculations
on the two representative structures in Figure \ref{cap:devices}.
Both devices have square QWs, with equal barriers on the top and bottom
interface. The first is typical of MODFETs and employs a donor layer
above the QW in order to populate it. This charge separation produces
an electric field between the two layers (across the barrier or spacer
layer) which can be approximated by \begin{equation}
E_{z}\approx\frac{en_{s}}{\epsilon_{0}\epsilon_{Si}}=-6\times10^{6}\textrm{ V/m},\label{eq:E-field}\end{equation}
 where $n_{s}=4\times10^{15}$ m$^{-2}$ is the density of electrons
in the 2DEG for Device 1, $e$ is the charge of an electron, and $\epsilon_{i}$
are the dielectric constants. We assume that \emph{the QW is populated
only by donor-layer electrons}, leaving an equal amount of positive
charge behind. The second structure is one that has been proposed
for use in a quantum computer device.\cite{ART-Friesen-PracticalQDQCinSi-2002}
It utilizes a near-lying, tunnel-coupled backgate 2DEG ($<30$ nm
away) together with Schottky top-gates to populate the QW selectively.
This situation also results in a strong electric field due to the
image potential on the back gate. For one qubit, this can be estimated
as \begin{equation}
E_{z}\approx\frac{e}{4\pi\epsilon_{0}\epsilon_{Si}d^{2}}=3\times10^{5}\textrm{ V/m},\end{equation}
 where $d=20$ nm is the distance from the QW to the back gate for
Device 2. Schottky top-gates and other device parameters can augment
or reduce this growth-direction electric field nominally up to the
breakdown field of silicon, $3\times10^{7}$ V/m, or the ionization
energy of the electron.\cite{ART-Jones/VASemi-2002} Indeed, this
field can actually be smaller than that due to the top-gates in certain
dot configurations.

The shift of the electron g-factor from its free-electron value $g_{0}=2.00232$
is one measure of SOC in a system. It is quite small in bulk silicon
and depends on the magnetic field direction in the (elliptical) conduction
band minima ($\Delta g_{\parallel}\approx-0.003$, $\Delta g_{\perp}\approx-0.004$).\cite{ART-Wilson/Feher-ESR3-1961}
However, it is difficult to reliably extract the SOC strength in a
2DEG from $\Delta g$. Many parameters (e.g., strain, barrier penetration,
Ge content ($g_{Ge}=1.4$), non-parabolicity of the band minima) influence
the magnitude and sign of $\Delta g$ and it may show considerable
sample dependence. The non-parabolicity effects are especially sensitive
to the electron density within the QW and can hide the magnitude of
SOC within a system.\cite{ART-Jiang/Yablonovich-2001}

In these silicon heterostructures, SOC is dominated by inversion asymmetry
within the device. The spin-orbit (SO) Hamiltonian to first order
in momentum is given in an arbitrary electrostatic potential $V$
by \[
H_{SO}=\frac{\hbar}{4m^{2}c^{2}}E_{z}\vec{\sigma}\cdot(\hat{z}\times\mathbf{p}),\]
 where $\sigma_{i}$ are the Pauli matrices. Note that the effective
magnetic field that acts on the spin is in the plane of the layer.
In Si heterostructures, the macroscopic fields, which do not average
out, are more important than the atomic electric fields. In the noncentrosymmetric
III-V materials such as GaAs, this is not necessarily the case and
the resulting Dresselhaus or \emph{bulk} inversion asymmetry fields
are usually dominant. The asymmetry considered here, due either to
an interface, charge distribution, or external potential, is usually
called Rashba or \emph{structural} inversion asymmetry.

The Rashba term comes directly from the SO Hamiltonian if we assume
one, dominant symmetry-breaking electric field in the structure and
average over a momentum state. In a QW, as we have pointed out above,
the electric field is in the growth ($z$) direction and thus the
$z$-component of the above dot-product is selected and we obtain
\begin{equation}
H_{R}^{2D}=\alpha(p_{x}\sigma_{y}-p_{y}\sigma_{x})\propto E_{z}(\vec{\sigma}\times\mathbf{p})_{z},\label{eq:Rashba-Eq}\end{equation}
 which is then the Rashba-Bychkov Hamiltonian.\cite{ART-Bychkov/Rashba-1984}

Strictly speaking, as de Andrade e Silva \emph{et. al.} point out
\cite{ART-AndradaESilva-SOinQWs}, the conduction-band-edge profile,
$E_{c}$, and the space charge separation (or applied electrostatic
field), $E_{z}$, contribute separately and sometimes dissimilarly
to the SOC. For example, the wavefunction discontinuity (band offset)
across a material-interface can cause Rashba spin-splitting itself.
However, in devices of the type considered here, the macroscopic field
should be the main contribution. These same authors have derived an
expression for $\alpha$ in the Kane model for GaAs. We have adapted
their work for Si, using a 5-parameter 8-band Kane model. This is
8 bands including spin, which means just the lowest conduction band
and the three highest valence bands. By calculating the breaking of
the degeneracy between the spin-up and spin-down states of the lowest
conduction band, we find \begin{equation}
\alpha=\frac{2PP_{z}\Delta_{d}}{\sqrt{2}\hbar E_{v1}E_{v2}}\left(\frac{1}{E_{v1}}+\frac{1}{E_{v2}}\right)e\left\langle E_{z}\right\rangle ,\label{eq:alphaSilva}\end{equation}
where we have taken the average of the electric field in the $z$-direction.
Here $P=\hbar\langle X|p_{x}|S\rangle/im$, $P_{z}=\hbar\langle Z|p_{z}|S\rangle/im$,
$\Delta_{d}=0.044$ eV is the spin-orbit splitting of the two highest
conduction bands, $E_{v1}=3.1$ eV is the direct gap of the strained
sample, and $E_{v2}=7$ eV is the gap between the conduction band
minimum and the lowest of the three valence bands. (These are the
5 parameters mentioned above.) $m$ is the bare electron mass. The
matrix elements that define $P$ and $P_{z}$ are to be taken between
the cell-periodic functions of the indicated symmetry at the position
of the conduction-band minimum. Unfortunately, these are not well
known in Si, since other bands contribute. We may note that $P$ and
$P_{z}$ are examples of momentum matrix elements that don't vary
too much in III-V materials and Ge,\cite{ART-Lawaetz-BandParams}
and are given approximately by $2mP_{(z)}^{2}/\hbar^{2}\approx22$
eV. With these values we find that for devices of type 1,\begin{equation}
\alpha\approx1.66\times10^{-6}\left\langle E_{z}\right\rangle \,\textrm{m/s.}\label{eq:alphaEfield}\end{equation}

Previous Kane models for GaAs involve matrix elements at $k=0$. Our
theory is new since it takes into account the proper symmetry of silicon
with it's minima well away from the zone center.

Wilamowski \emph{et. al.} \cite{ART-Wilamowsky/Rossler-2002}, using
conduction electron spin resonance (CESR), have measured $\alpha\approx5.94$
m/s ($\alpha e/\hbar=0.55/\sqrt{2}\times10^{-12}$ eV$\cdot$cm in
their units) %
\footnote{Note that \cite{ART-Wilamowsky/Rossler-2002} uses the wrong expression
for the Fermi wavevector, failing to include the valley degeneracy,
so their data analysis is off by a factor of sqrt(2).%
} in a Si$_{0.75}$Ge$_{0.25}$/Si/Si$_{0.75}$Ge$_{0.25}$ QW where
the strained-silicon layer was roughly $12-20$ nm. Carrier concentrations
were $n_{s}\approx4\times10^{15}\, m^{-2}$. These numbers correspond
to our Device 1 parameters. Our equations then give, using Eq. \ref{eq:alphaEfield},\begin{equation}
\alpha^{D1}\approx5.1\textrm{ m/s for $\left\langle E_{z}\right\rangle =3\times10^{6}$ V/m}.\label{eq:alhpaDevice1}\end{equation}
 Theory compares in order of magnitude and we believe that our estimation
has some utility as a guide for device design.

For Device 2, $\alpha^{D2}\approx0.25\textrm{ m/s}$ for $\left\langle E_{z}\right\rangle =1.5\times10^{5}$
V/m. This device remains to be built. For Device 1, we can also predict
the zero magnetic field spin-splitting in a silicon 2DEG using $\left|\epsilon_{+}-\epsilon_{-}\right|\leq2\alpha p_{F},$\[
2\alpha p_{F}=2\alpha\hbar\sqrt{\frac{4\pi n_{s}}{4}}\approx0.75\,\mu\textrm{eV},\]
 where $4$ is the degeneracy factor in silicon (spin+valley). This
has not yet been directly measured to our knowledge. Taking a Zeeman
splitting of $g\mu_{B}B=0.75\mu\textrm{eV}$ with a g-factor of 2,
this implies an internal, in-plane, effective magnetic field---the
so-called \emph{Rashba field}---of roughly 62 Gauss, which is the
direct result of SOC in the silicon 2DEG of Device 1. 

Let's consider the relevance of our silicon SOC results to QC and
spintronics. We note that the magnitude of the Rashba coefficient
is much smaller than for GaAs. For the same electric field and 2DEG
density as Device 1, a similar $\vec{k}\cdot\vec{p}$ theory for GaAs
arrives at $\alpha_{GaAs}\approx230$ m/s.\cite{ART-AndradaESilva-SOinQWs}
But GaAs itself is not a high Rashba III-V semiconductor and is thought
to be be dominated by Dresselhaus SOC ($\beta_{GaAs}\approx1000$
m/s).\cite{ART-Khaetskii-SpinRelaxationInDots-1999} InAs-based heterojuctions,
for example, may have orders of magnitude higher Rashba values.\cite{ART-Cui-InAsRashba}
This means that SOC effects in silicon devices will be much smaller,
including decoherence and gating errors that are SOC-based.

\section{\textbf{Spin relaxation}}

We wish to consider the combined effects of the SO Hamiltonian

\[
H_{R}^{2D}=\alpha(p_{x}\sigma_{y}-p_{y}\sigma_{x})\]
 and the scattering Hamiltonian. The scattering may be from phonons
or from static disorder. We take the semiclassical approach, in which
the effect of scattering is to cause transitions at random intervals
from one wavepacket centered at $\vec{p}$ with $\varepsilon_{\vec{p}}=\varepsilon_{F}$
to another centered at $\vec{p}^{\prime}$ with $\varepsilon_{\vec{p}^{\prime}}=\varepsilon_{F},$
where $\varepsilon_{F}$ is the Fermi energy. This corresponds to
a random switching in the direction of the effective magnetic field
that acts on the spin degree of freedom. This is the D'yakonov-Perel'
mechanism of spin relaxation.\cite{ART-Dyakonov/Perel-3DEG-1972,ART-Dyakonov/Kachorovskii-1986}
The measured quantity in the continuous-wave experiments carried out
on 2DEGs is $T_{2},$ the transverse relaxation time, while pulsed
experiments can also measure $T_{1},$ the longitudinal relaxation
time. For our purposes, a density matrix approach is the natural one,
since we will eventually want to perform an ensemble average over
all possible scattering sequences. Since the physical model of spins
in a random time-dependent magnetic field is the same as that for
relaxation of nuclear spins in liquids, the Redfield technique may
be used.%
\begin{figure}
\begin{center}\includegraphics[%
  scale=0.4]{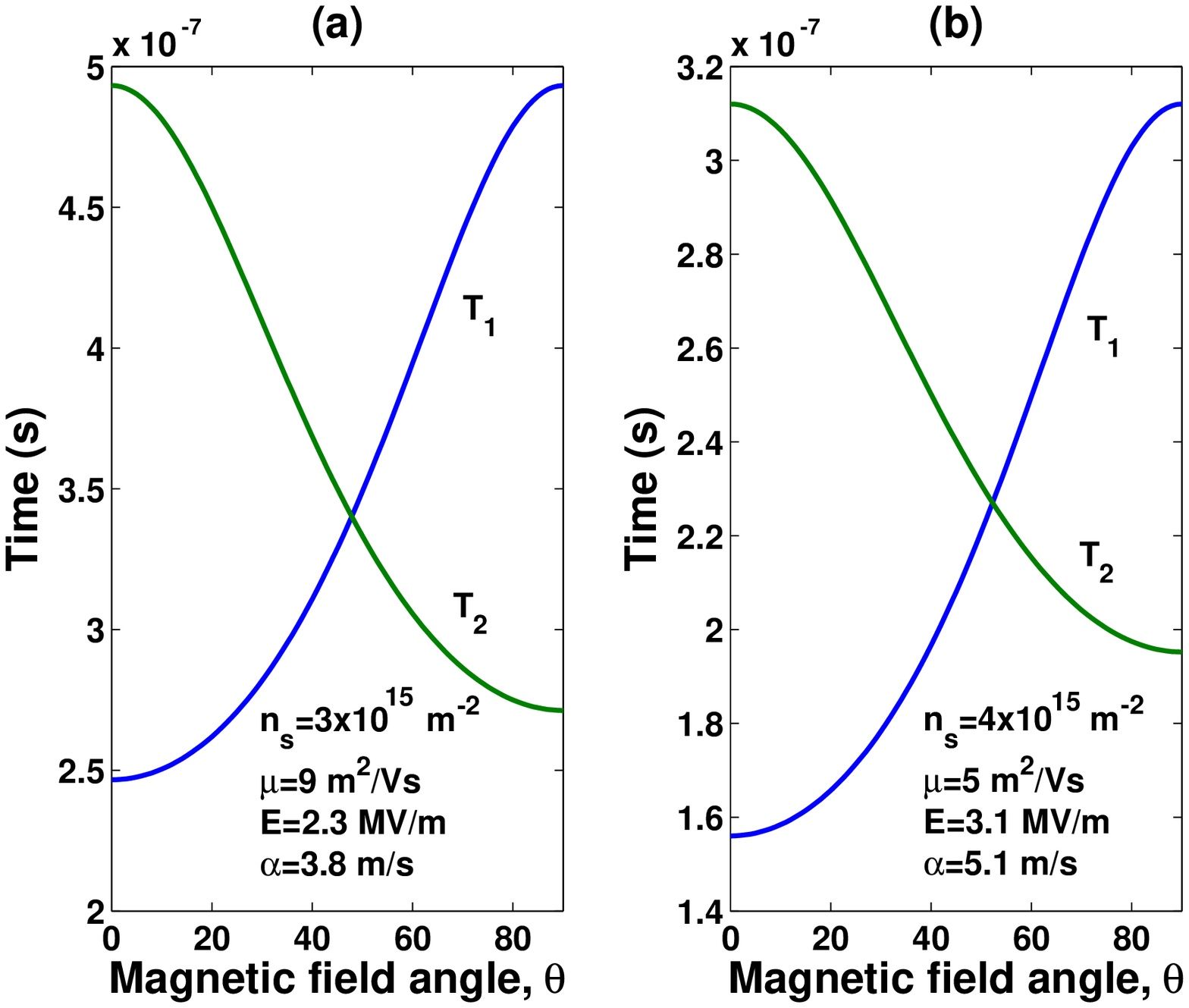}\end{center}

\begin{center}\includegraphics[%
  scale=0.4]{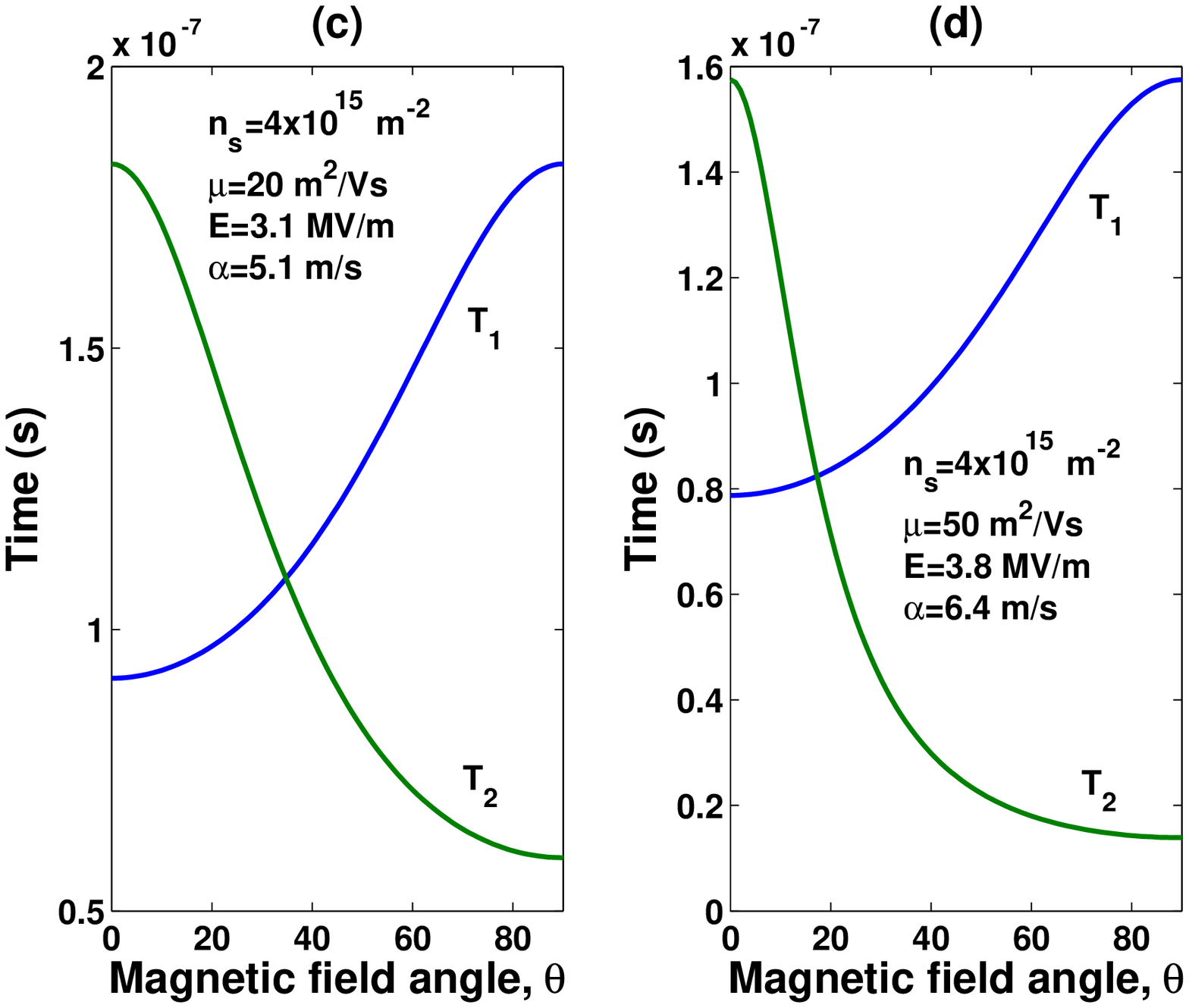}\end{center}

\caption{ESR linewidths in Gauss as a function of static magnetic field direction
(where $\theta=0$ is perpendicular to the 2DEG plane) for specific
values of 2DEG density and Rashba asymmetry. The quantum well is assumed
to be completely donor-layer populated and as such, $\alpha$ is calculated
directly with Eq. \ref{eq:alphaEfield} as a function of the 2DEG
density.\label{cap:Linewidth}}
\end{figure}

\begin{figure}
\begin{center}\includegraphics[%
  scale=0.7]{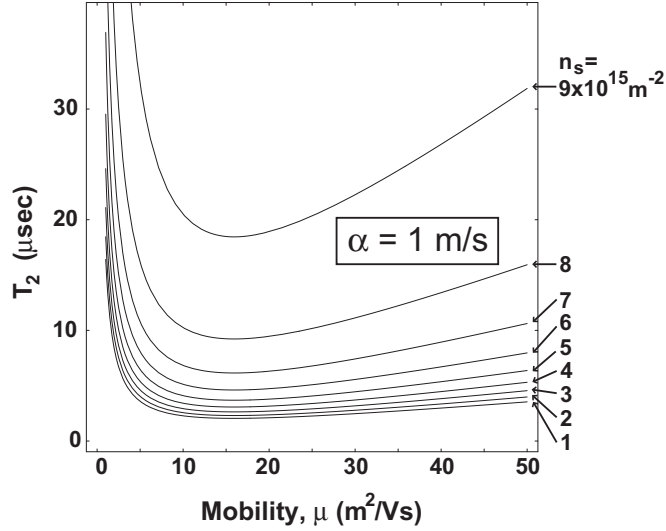}\end{center}

\caption{ESR linewidth lifetime $T_{2}$ from Eq. \ref{eq:T2} for constant
asymmetry coefficient, $\alpha=1$ m/s, as a function of 2DEG mobility,
$\mu$, and density, $n_{s}$. For donor-layer populated quantum wells,
divide the times listed by $\alpha^{2}$: $T_{2}(\alpha)=T_{2}(\alpha=1)/\alpha^{2}$.
The magnetic field is assumed to be $B=0.33$ Tesla, perpendicular
to the plane of the 2DEG.\label{cap:Times}}
\end{figure}

We outline the calculation only briefly, since the details are parallel
to the discussion in standard texts.\cite{BOOK-Slichter-MagneticResonance}
The $2\times2$ density matrix $\rho$ allows us to compute the expectation
values of the spin by $\left\langle \sigma_{i}\right\rangle =Tr~\left\{ \sigma_{i}\rho\right\} .$
For a single system described by the Hamiltonian $H,$ we have the
equation of motion $d\rho/dt=\frac{i}{\hbar}\left[\rho,H\right].$
In the case of \begin{equation}
H=H_{0}+H_{1}(t),\end{equation}
where $H_{1}$ is small, it is convenient to go to the interaction
representation\begin{equation}
\rho^{_{int}}=\exp(-iH_{0}t/\hbar)\rho\exp(iH_{0}t/\hbar),\end{equation}
and then we get\begin{equation}
\frac{d\rho^{_{int}}}{dt}=\frac{i}{\hbar}\left[\rho^{_{int}},H_{1}^{int}(t)\right],\end{equation}
where\begin{equation}
H_{1}^{int}(t)=\exp(iH_{0}t/\hbar)H_{1}\exp(-iH_{0}t/\hbar).\end{equation}
This equation can be integrated to give\begin{equation}
\rho^{_{int}}(t)=\rho^{_{int}}(0)+\frac{i}{\hbar}\int_{0}^{t}\left[\rho^{_{int}}(t^{\prime}),H_{1}^{int}(t^{\prime})\right]dt^{\prime},\end{equation}
and this can be solved interatively, which in second order gives\begin{eqnarray}
\frac{d\rho^{_{int}}(t)}{dt} & = & \frac{i}{\hbar}\left[\rho^{_{int}}(0),H_{1}^{int}(t)\right]\\
 & + & \left(\frac{i}{\hbar}\right)^{2}\int_{0}^{t}dt^{\prime}\left[\left[\rho^{_{int}}(0),H_{1}^{int}(t^{^{\prime}})\right],H_{1}^{int}(t)\right].\nonumber \end{eqnarray}

For example, let the steady field be in the $z$-direction, so that
$H_{0}=\hbar\omega_{c}\sigma_{z}/2.$ The fluctuating field $\left[H_{_{1}}^{int}(t)\right]_{ss^{\prime\prime}}=\sum_{i=x,y}h^{i}(t)\sigma_{ss^{\prime\prime}}^{i}$
is in the transverse direction. The first order matrix element vanishes
and we are left, in second order, with \begin{eqnarray}
\left(\frac{i}{\hbar}\right)^{-2}\left(\frac{d\rho_{ss^{\prime}}^{int}}{dt}\right)_{2} & = & \sum_{s^{^{\prime\prime}}s^{^{\prime\prime\prime}}}\rho_{ss^{\prime\prime}}(0)\left[H_{_{1}}^{int}(t^{\prime})\right]_{s^{\prime\prime}s^{\prime\prime\prime}}\left[H_{_{1}}^{int}(t)\right]_{s^{\prime\prime\prime}s^{\prime}}e^{i(s^{\prime\prime}-s^{\prime\prime\prime})t^{\prime}}e^{i(s^{\prime\prime\prime}-s^{\prime})t}\label{eq:pert}\\
 &  & +\sum_{s^{\prime\prime}s^{\prime\prime\prime}}\left[H_{_{1}}^{int}(t)\right]_{ss^{\prime\prime}}e^{i(s-s^{\prime\prime})t}\left[H_{_{1}}^{int}(t)\right]_{s^{\prime\prime}s^{\prime\prime\prime}}e^{i(s^{\prime\prime}-s^{\prime\prime\prime})t^{\prime}}\rho_{s^{\prime\prime\prime}s^{\prime}}(0)\nonumber \\
 &  & -\sum_{s^{\prime\prime}s^{\prime\prime\prime}}\left[H_{_{1}}^{int}(t^{\prime})\right]_{ss^{\prime\prime}}e^{i(s-s^{\prime\prime})t^{\prime}}\rho_{s^{\prime\prime}s^{\prime\prime\prime}}^{_{int}}(0)\left[H_{_{1}}^{int}(t)\right]_{s^{\prime\prime\prime}s^{\prime}}e^{i(s^{\prime\prime\prime}-s^{\prime})t}\nonumber \\
 &  & -\sum_{s^{\prime\prime}s^{\prime\prime\prime}}\left[H_{_{1}}^{int}(t)\right]_{ss^{\prime\prime}}e^{i(s-s^{\prime\prime})t}\rho_{s^{\prime\prime}s^{\prime\prime\prime}}^{_{int}}(0)e^{i(s^{\prime\prime\prime}-s^{\prime})t^{\prime}}\left[H_{_{1}}^{int}(t^{\prime})\right]_{s^{\prime\prime\prime}s^{\prime}},\nonumber \end{eqnarray}
where we define $e^{st}=e^{\epsilon_{s}t}.$ Averaging the random
field, we find\begin{eqnarray*}
\overline{\left[H_{_{1}}^{int}(t)\right]_{ss^{\prime\prime}}\left[H_{_{1}}^{int}(t^{\prime})\right]_{s^{\prime\prime}s^{\prime\prime\prime}}} & = & \sum_{ij}\overline{h^{i}(t)h^{j}(t^{\prime})}\sigma_{ss^{\prime\prime}}^{i}\sigma_{s^{\prime\prime}s^{\prime\prime\prime}}^{j}\\
 & = & \sum_{i=x,y,z}\chi^{i}(t-t^{\prime})\sigma_{ss^{\prime\prime}}^{i}\sigma_{s^{\prime\prime}s^{\prime\prime\prime}}^{i},\end{eqnarray*}
where $\tau=t-t^{\prime}$ and we define $\chi^{i}(\tau)=$ $\overline{h^{i}(t)h^{i}(t^{\prime})}$.
We substitute this expression into Eq. \ref{eq:pert} and do the matrix
algebra. Eq. \ref{eq:pert} can then be integrated in the limit where
$t$ is large. We neglect the oscillating terms, which then yield
\[
\left(\frac{i}{\hbar}\right)^{-2}\left\langle \frac{d\rho_{++}}{dt}\right\rangle =\left[\chi^{x}(\omega_{c})+\chi^{y}(\omega_{c})\right]\left[\rho_{++}(0)-\rho_{--}(0)\right]\]
and\[
\left(\frac{i}{\hbar}\right)^{-2}\left\langle \frac{d\rho_{+-}}{dt}\right\rangle =\rho_{+-}(0)\left[\chi^{x}(\omega_{c})+\chi^{y}(\omega_{c})+2\chi^{z}(0)\right]\]
To obtain the relaxation times we must consider the equation for the
spin: \begin{eqnarray}
\frac{d\left\langle \sigma^{z}\right\rangle }{dt} & = & \frac{d\left[Tr\left(\sigma^{z}\rho\right)\right]}{dt}=\frac{d}{dt}(\rho_{++}-\rho_{--})\nonumber \\
 & = & 2\left(\frac{i}{\hbar}\right)^{2}\left[\chi^{x}(\omega_{c})+\chi^{y}(\omega_{c})\right]\left[\rho_{++}(0)-\rho_{--}(0)\right]\nonumber \\
 & = & -2\left(\frac{1}{\hbar}\right)^{2}\left[\chi^{x}(\omega_{c})+\chi^{y}(\omega_{c})\right]\left\langle \sigma_{z}\right\rangle \end{eqnarray}
and so \[
1/T_{1}=2\left[\chi^{x}(\omega_{c})+\chi^{y}(\omega_{c})\right]/\hbar^{2}.\]
Also

\begin{eqnarray*}
\frac{d\left\langle \sigma^{x}\right\rangle }{dt} & = & \frac{dTr\left(\sigma^{x}\rho\right)}{dt}=\frac{d}{dt}(\rho_{+-}+\rho_{-+})\\
 & = & -\left[\rho_{+-}(0)+\rho_{+-}(0)\right]\left[\chi^{x}(\omega_{c})+\chi^{y}(\omega_{c})+2\chi^{z}(0)\right]/\hbar^{2}\\
 & = & -\left[\chi^{x}(\omega_{c})+\chi^{y}(\omega_{c})+2\chi^{z}(0)\right]\left\langle \sigma_{x}\right\rangle /\hbar^{2},\end{eqnarray*}
which gives\[
1/T_{2}=\left[\chi^{x}(\omega_{c})+\chi^{y}(\omega_{c})+2\chi^{z}(0)\right]/\hbar^{2}.\]
This gives a relation\[
1/T_{2}=1/2T_{1}+2\chi^{z}(0)/\hbar^{2}.\]
We now wish to specialize to the case of a DP mechanism in a 2DEG.
The main point is that the static field may be in any direction, while
the fluctuationg field is in the plane. Consider first the special
case that the static field is along $\widehat{z}.$ Then \[
1/T_{2}=1/2T_{1}=\left[\chi^{x}(\omega_{c})+\chi^{y}(\omega_{c})\right]/\hbar^{2}.\,\,\]
Now consider a general direction, say $\vec{B}$ along the direction
$B_{x}\widehat{x}+B_{z}\widehat{z}=\sin\theta\widehat{x}+\cos\theta\widehat{z}$,
so that $\theta$ is the angle to the normal. Then the longitudinal
fluctuations $\chi^{\parallel}$, which are quadratic in the field,
are proportional to $\sin^{2}\theta$ and the transverse ones $\chi^{\perp}$to
$\cos^{2}\theta.$ Thus\begin{equation}
1/T_{1}(\theta)=2\left[\cos^{2}\theta\chi^{x}(\omega_{L})+\chi^{y}(\omega_{L})\right]/\hbar^{2}\label{eq:T1}\end{equation}
 while\begin{equation}
1/T_{2}(\theta)=\left[\cos^{2}\theta\chi^{x}(\omega_{L})+\chi^{y}(\omega_{L})+2\sin^{2}\theta\chi^{x}(0)\right]/\hbar^{2}.\label{eq:T2}\end{equation}
For the DP mechanism in a 2DEG the random field is constant in magnitude,
but random in direction. The statistics of this field are Poisson:
namely that if the $x$-component the field at time $t=0$ is $h^{x},$
then the chance of it remaining at $h^{x}$ decays as $\exp(-t/\tau_{p})$.
Hence \[
\overline{h^{i}(0)h^{i}(\tau_{p})}=\left\langle h_{x}^{2}\right\rangle e^{-t/\tau}=\frac{1}{2}\alpha^{2}p_{F}^{2}e^{-t/\tau_{p}},\]
and\[
\chi^{x,y}(\omega)=\frac{\alpha^{2}p_{F}^{2}\tau_{p}^{2}}{1+\omega^{2}\tau_{p}^{2}}.\]
In these formulas, $\tau_{p}$ is the momentum relaxation time. Note
that these formulas assume s-wave scattering.

The zero-frequency limit of these formulas agrees with the recent
results of Burkov and MacDonald \cite{Burkov/MacDonald-2DEGRashba-2003}
{[}see their Eq.(17){]} . They do not agree with the formulas in Wilamowski
\emph{et. al.} \cite{ART-Wilamowsky/Rossler-2002}, {[}see, e.g.,
their Eq.(3){]} who state that the relaxation from the DP mechanism
should vanish when $\theta=0$. This is not consistent with our results.

The DP mechanism has the nice feature that it is relatively easy to
isolate experimentally. It is strongly anisotropic in the direction
of the applied field compared to other mechanisms. To illustrate this
we plot the ESR linewidths as a function of field angle in Fig. \ref{cap:Linewidth}.
What is most striking is the opposite dependence on angle for the
rates $1/T_{1}$ and $1/T_{2}$, with $1/T_{2}$ maximized when the
field is in the plane of the 2DEG, while $1/T_{1}$ is maximized when
the field is perpendicular to the plane of the 2DEG. Physically, this
comes from the fact that the electric field is perpendicular, so that
the fluctuations of the effective magnetic field are in the plane.
Longitudinal relaxation ($T_{1}$) is due to fluctuations perpendicular
to the steady field, while transverse relaxation ($T_{2}$) is due
to fluctuations both perpendicular and parallel to the steady field.
This mechanism has the characteristic that the change in $1/T_{1}$
as the field is rotated through 90 degrees is always a factor of two.
The change in $1/T_{2}$ is frequency- and lifetime-dependent, with
the anisotropy increasing as the mobility increases. 

The DP relaxation also has the counter-intuitive inverse dependence
of the spin relaxation time on the momentum relaxation time: $1/T_{1,2}\propto\tau_{p}$,
for small $\tau_{p}$ (or zero field), typical for motional narrowing.
We plot the dependence of $T_{2}$ on the mobility in Fig. \ref{cap:Times}.
At high mobilities and high frequencies $\omega>>1/\tau_{p}$, we
find $1/T_{1,2}\propto1/\tau_{p}$. %
\begin{table}
\begin{center}\begin{tabular}{c|c|c}
Source&
Linewidth&
Anisotropy\tabularnewline
\hline 
\begin{tabular}{c}
Ref. \cite{ART-Wilamowsky/Rossler-2002} \tabularnewline
5-30 K\tabularnewline
$\mu\sim20$ m$^{2}$/V-s\tabularnewline
$n_{s}\sim4\times10^{15}$ m$^{-2}$\tabularnewline
CW-ESR\tabularnewline
\textbf{donor} populated\tabularnewline
\end{tabular}&
\begin{tabular}{c}
Exp.:\tabularnewline
$T_{2}^{B\parallel z}=420$ ns (0.15 G)\tabularnewline
$T_{2}^{B\perp z}=140$ ns (0.45 G)\tabularnewline
Pred.:\tabularnewline
$T_{2}^{B\parallel z}=$191 ns\tabularnewline
$T_{2}^{B\perp z}=$60 ns\tabularnewline
\end{tabular}&
\begin{tabular}{c}
 $T_{2}^{B\parallel z}$/$T_{2}^{B\perp z}$= 3\tabularnewline
\tabularnewline
\tabularnewline
$T_{2}^{B\parallel z}$/$T_{2}^{B\perp z}$= 3.2\tabularnewline
$T_{1}^{B\parallel z}$/$T_{1}^{B\perp z}$= 0.5 \tabularnewline
\tabularnewline
\end{tabular}\tabularnewline
\hline
\begin{tabular}{c}
Ref. \cite{ART-Tryshkin/Lyon-ESRof2DEGs-2003} \tabularnewline
5 K\tabularnewline
$\mu\sim9$ m$^{2}$/V-s\tabularnewline
$n_{s}\sim3\times10^{15}$m$^{-2}$\tabularnewline
Pulsed-ESR\tabularnewline
\textbf{light} populated\tabularnewline
\end{tabular}&
\begin{tabular}{c}
Exp.:\tabularnewline
$T_{2}^{B\parallel z}=$3 $\mu$s\tabularnewline
$T_{2}^{B\perp z}=$0.24 $\mu$s\tabularnewline
$T_{1}^{B\parallel z}=$2 $\mu$s\tabularnewline
$T_{1}^{B\perp z}=$3 $\mu$s\tabularnewline
Pred.:\tabularnewline
$T_{2}^{B\parallel z}=$502 ns\tabularnewline
$T_{2}^{B\perp z}=$272 ns\tabularnewline
$T_{1}^{B\parallel z}=$251 ns\tabularnewline
$T_{1}^{B\perp z}=$502 ns\tabularnewline
\end{tabular}&
\begin{tabular}{c}
 $T_{2}^{B\parallel z}$/$T_{2}^{B\perp z}$= 12.5\tabularnewline
$T_{1}^{B\parallel z}$/$T_{1}^{B\perp z}$=0.67 \tabularnewline
\tabularnewline
$T_{2}^{B\parallel z}$/$T_{2}^{B\perp z}$= 1.8\tabularnewline
$T_{1}^{B\parallel z}$/$T_{1}^{B\perp z}$=0.5 \tabularnewline
\tabularnewline
\end{tabular}\tabularnewline
\hline
\begin{tabular}{c}
Ref. \cite{ART-Jantsch/Schaffler-2002} \tabularnewline
\tabularnewline
$\mu\sim10$ m$^{2}$/V-s\tabularnewline
$n_{s}\sim3\times10^{15}$ m$^{-2}$\tabularnewline
CW-ESR\tabularnewline
\textbf{light/gate} populated\tabularnewline
\end{tabular}&
\begin{tabular}{c}
Exp.:\tabularnewline
$T_{2}^{B\parallel z}=$12 $\mu$s\tabularnewline
$T_{2}^{B\perp z}=$500 ns\tabularnewline
Pred.:\tabularnewline
$T_{2}^{B\parallel z}=$480 ns\tabularnewline
$T_{2}^{B\perp z}=$193 ns\tabularnewline
\end{tabular}&
\begin{tabular}{c}
\tabularnewline
$T_{2}^{B\parallel z}$/$T_{2}^{B\perp z}$= 24\tabularnewline
\tabularnewline
$T_{2}^{B\parallel z}$/$T_{2}^{B\perp z}$= 1.9\tabularnewline
$T_{1}^{B\parallel z}$/$T_{1}^{B\perp z}$= 0.5 \tabularnewline
\tabularnewline
\end{tabular}\tabularnewline
\hline
\begin{tabular}{c}
Ref. \cite{ART-Truitt-Si2DEGs} \tabularnewline
\tabularnewline
$\mu\sim5$ m$^{2}$/V-s\tabularnewline
$n_{s}\sim4\times10^{15}$ m$^{-2}$\tabularnewline
CW-ESR\tabularnewline
\textbf{donor} populated\tabularnewline
\end{tabular}&
\begin{tabular}{c}
Exp.:\tabularnewline
$T_{2}^{B\parallel z}=50$ ns (0.13 G) \tabularnewline
$T_{2}^{B\perp z}=30$ ns (0.215 G)\tabularnewline
Pred.:\tabularnewline
$T_{2}^{B\parallel z}=$31 ns\tabularnewline
$T_{2}^{B\perp z}=$20 ns\tabularnewline
\end{tabular}&
\begin{tabular}{c}
\tabularnewline
$T_{2}^{B\parallel z}$/$T_{2}^{B\perp z}$= 1.65\tabularnewline
\tabularnewline
$T_{2}^{B\parallel z}$/$T_{2}^{B\perp z}$= 1.6\tabularnewline
$T_{1}^{B\parallel z}$/$T_{1}^{B\perp z}$= 0.5 \tabularnewline
\tabularnewline
\end{tabular}\tabularnewline
\hline
\begin{tabular}{c}
Ref. \cite{ART-Graeff-2DEGESRSi} \tabularnewline
4.2 K\tabularnewline
$\mu\sim9$ m$^{2}$/V-s\tabularnewline
$n_{s}\sim4\times10^{15}$ m$^{-2}$\tabularnewline
ED-ESR\tabularnewline
\textbf{donor} populated\tabularnewline
\end{tabular}&
\begin{tabular}{c}
Exp.:\tabularnewline
$T_{2}^{B\parallel z}=105$ ns (0.6 G)\tabularnewline
$T_{2}^{B\perp z}=49$ ns (1.3 G)\tabularnewline
Pred.:\tabularnewline
$T_{2}^{B\parallel z}=$212 ns\tabularnewline
$T_{2}^{B\perp z}=$115 ns\tabularnewline
\end{tabular}&
\begin{tabular}{c}
\tabularnewline
$T_{2}^{B\parallel z}$/$T_{2}^{B\perp z}$= 2.2\tabularnewline
\tabularnewline
$T_{2}^{B\parallel z}$/$T_{2}^{B\perp z}$= 1.8\tabularnewline
$T_{1}^{B\parallel z}$/$T_{1}^{B\perp z}$= 0.5 \tabularnewline
\tabularnewline
\end{tabular}\tabularnewline
\hline
\end{tabular}\end{center}

\caption{We calculate relaxation times assuming that the QW is completely
populated by the donor layer. Accurate analysis is made difficult
due to lack of precise values for mobility and density, which are
often not measured directly (or reported) for the specific sample
addressed with ESR. The anisotropy does not depend on the Rashba coefficient.
Note also that there is some disagreement in the literature as to
how to convert from linewidth to a relaxation time, we use the equations
derived by Poole in Ref. \cite{BOOK-Poole}, $T_{2}=2\hbar/\left(\sqrt{3}g\mu_{B}\Delta H_{pp}^{0}\right),$
but others may differ by a factor of up to $2\pi$. \label{cap:Samples}}
\end{table}

\section{\textbf{Discussion}}

We have calculated the transverse and longitudinal relaxation times
of a silicon 2DEG in an arbitrary static magnetic field. To test our
calculations, we compare them to known ESR data.\cite{ART-Jantsch/Schaffler-2002,ART-Tryshkin/Lyon-ESRof2DEGs-2003,ART-Wilamowsky/Rossler-2002,ART-Truitt-Si2DEGs,ART-Graeff-2DEGESRSi}
We limit ourselves to low temperatures, $\epsilon_{F}\sim10-15$ K,
and realistic material parameters for state-of-the-art heterostructures.

The most robust prediction of the theory is the anisotropy, particularly
that of $T_{1}$, which is completely independent of all parameters.
The only measurement, in Ref. {[}3{]}, gives satisfactory agreement
for $T_{1}: T_{1}^{B\parallel z}/T_{1}^{B\perp z}=0.67$ as opposed
to the prediction 0.5. Furthermore, the anisotropy of $1/T_{2}$ goes
in the opposite direction, as it should. The magnitude of this anisotropy
is measured to be $T_{2}^{B\parallel z}/T_{2}^{B\perp z}=12.5$ which
is about a factor of six larger than the theory predicts for the quoted
mobility. The relaxation times, as far as can be determined by the
range set by the uncertainty in silicon band parameters, are in rough
agreement with what one gets from estimates assuming that this well
is a device of type 1. The anisotropy of the 2DEG ESR linewidth is
independent of $\alpha$ and is indeed only dependent on one free
variable: the momentum relaxation time $\tau_{p}$ which we assume
is directly proportional to the mobility. The magnitude of the relaxation
time, on the other hand, is set by the Rashba coefficient together
with the Fermi momentum. 

Table \ref{cap:Samples} details results for four other samples as
well, also nominally of type 1 for which measurements of $T_{2}$
have been performed. Agreement is good for the sample of Ref. \cite{ART-Wilamowsky/Rossler-2002},
particularly for the anisotropy of $T_{2}$. This is a donor-layer-populated
sample measured with CW-ESR. Ref. \cite{ART-Graeff-2DEGESRSi}, measured
via electrically-detected ESR (ED-ESR), also is in good agreement
with the anisotropy predicted by the theory. Comparison with the sample
of Ref. \cite{ART-Truitt-Si2DEGs} also seems to be very good. This
is an IBM 2DEG with a density of roughly $4\times10^{15}$ m$^{-2}$
and a quantum well thickness of 10 nm, fully donor-layer populated.
Agreement is considerably less good for the sample in Ref. \cite{ART-Jantsch/Schaffler-2002}.
Here the mobility is not well-known, and the sample is partially populated
by illumination. So we are lead to believe that there is some difference
between the two sets of samples that causes one set to have a larger
$T_{2}$ anisotropy than our theory predicts, even for very similar
material paramters (density and mobility). Further experimental work
along the lines of measuring the density, mobility, $T_{1}$, and
$T_{2}$ on the same samples is needed. 

The magnitude of the predicted relaxation times is in general well
predicted by the theory, but there is a range of error. The position
of the ionized centers which populate the well is important for the
calculation of the electric field which is thus hard to characterize.
Photoelectrons created with light at the bandgap energy may also have
symmetry changing effects. This may explain why the Rashba coefficient
derived from varying the 2DEG density by light in Ref. \cite{ART-Wilamowsky/Rossler-2002}
appears to be independent of density. It is also important to point
out that parallel conductivity (current paths through both the 2DEG
and the donor-layer for example) is a common problem in today's SiGe
quantum wells and may effect the transport measurements of density
and especially mobility making comparison with theory difficult.

Other mechanisms may become important as we leave the parameter range
considered in this paper. Electron-electron collisions which do not
greatly affect the mobility at low temperatures may start to contribute
at higher temperatures and mobilities, as they appear to do in GaAs
quantum wells.\cite{ART-Glazov/Harley-2003} These collisions will
also relax the spin, but the relation between momentum relaxation
and spin relaxation is not expected to be the same as for the elastic
collisions considered here. At higher magnetic fields, the cyclotron
motion of the electrons is important. In the semiclassical picture,
when $\omega_{c}\tau_{p}\geq1$, the average value of the momentum
perpendicular to the magnetic field shrinks, reducing or even eliminating
DP spin relaxation, as has been considered for III-V semiconductors
in the fixed magnetic field case.\cite{BOOK-Optical-Orientation,ART-Ivchenko-cyclo}
This effect may become important at high mobilities and would be dependent
on magnetic field angle, increasing the anisotropy predicted here
while also increasing the relaxation times. Quantum effects may become
important in this regime however. The wave vector dependence of the
conduction band electron g-factor may also lead to relaxation, as
has been pointed out recently,\cite{ART-Bronold-2deg} but hasn't
been considered here. Finally, the addition of details related to
the presence of two conduction-band valleys may differentiate further
the case of Si from that of GaAs. Golub and Ivchenko \cite{ART-Golub/Ivchenko-SiGeSplitting-2003}
have considered spin relaxation in symmetrical ($\alpha$=0) SiGe
QWs, where valley domains (even or odd monolayer regions of the QW)
may have influence over spin dynamics. Random spin-orbit coupling
due to variations in the donor layer charge distribution may also
be important in symmetric quantum wells.\cite{ART-Sherman-APLRandomSO,ART-Sherman-PRB}

Long spin relaxation times on the order of hundreds of nanoseconds
to microseconds, found in presently available SiGe quantum wells,
hold great promise for both quantum information processing and spintronics.
Our results demonstrate that decreasing the reflection asymmetry within
the device will appreciably decrease the Rashba coefficient and the
consequent spin relaxation at low temperatures. This can be achieved
by a symmetric doping profile or using an external electric field
to cancel out the field of the ions. They further show that the anisotropy
of the ESR linewidth as a function of angle may be a good indicator
of 2DEG quality (mobility) independent of transport measurements.
As higher mobility and more exotic SiGe heterostructures are grown
and characterized, new physics may emerge.

\section*{Acknowledgements}

The authors would like to acknowledge useful conversations with Mark
Friesen and Jim Truitt and with the rest of the Wisconsin-Madison
solid-state quantum computing group.%
\footnote{http://qc.physics.wisc.edu/%
} This work has been supported by the Army Research Office through
ARDA and the NSF QUBIC program NSF-ITR-0130400. \bibliographystyle{apsrev}
\bibliography{/Volumes/cgtWork/Research/Bib/ctqc}

\end{document}